\documentclass[twocolumn,prl]{revtex4}
\usepackage{graphicx}
\usepackage{amssymb,amsmath,bm}
\newcommand{\be}{\begin{equation}}
\newcommand{\ee}{\end{equation}}
\newcommand{\bea}{\begin{eqnarray}}
\newcommand{\eea}{\end{eqnarray}}

\newcommand{\lp}{\left(}
\newcommand{\rp}{\right)}

\renewcommand{\phi}{\varphi}
\renewcommand{\epsilon}{\varepsilon}
\renewcommand{\vec}[1]{{\bf #1}}

\renewcommand{\Im}{\mathop{\rm Im}\nolimits}

\begin{document}
\title{Atomic Collapse
and Quasi-Rydberg States in Graphene}
\author{
A. V. Shytov,${}^1$ M. I. Katsnelson,${}^2$ L. S. Levitov${}^3$}
\affiliation{
${}^1$ 
Brookhaven National Laboratory, Upton, 
New York 11973-5000\\
${}^2$ Radboud University of Nijmegen, Toernooiveld 1
6525 ED Nijmegen,
The Netherlands, \\
${}^3$ Department of Physics,
 Massachusetts Institute of Technology, 77 Massachusetts Ave,
 Cambridge, MA 02139
}


\begin{abstract}
Charge impurities in graphene can host
an infinite family of Rydberg-like resonance states 
of massless Dirac particles.
These states, appearing for supercritical charge, 
are described by Bohr-Sommerfeld quantization
of {\it collapsing} classical trajectories that descend on point charge,
in analogy to Rydberg states relation with planetary orbits.
We argue that divalent and trivalent charge impurities in graphene 
is an ideal system
for realization of this atomic collapse regime.
Strong coupling of these states
to the Dirac continuum via Klein tunneling leads to striking resonance effects 
with direct signatures in transport,
local properties, and enhancement of the Kondo effect.
\end{abstract}
\maketitle

The discovery of massless Dirac excitations in graphene\,\cite{reviewGK}
triggered new interest in 
solid-state realization of quantum electrodynamics (QED)\,\cite{Gonzalez94,Katsnelson06b,KatsnelsonSSC}.
Transport phenomena in this system\,\cite{Cheianov06,Cheianov07} can be used to probe 
classic concepts of QED, such as chiral dynamics\,\cite{Greiner_book},
flavor degrees of freedom\,\cite{Rycerz07} 
and particle/hole coexistence\,\cite{Abanin07}. 
Here we demonstrate that graphene opens a way to investigate 
in the laboratory a fundamental 
quantum relativistic phenomenon, that is, atomic collapse
in a strong Coulomb electric field\,\cite{Zeldovich,Greiner_book}, 
long sought for
but still inaccessible in high-energy experiments\,\cite{SLAC}.

\begin{figure}
\includegraphics[height=1.35in]{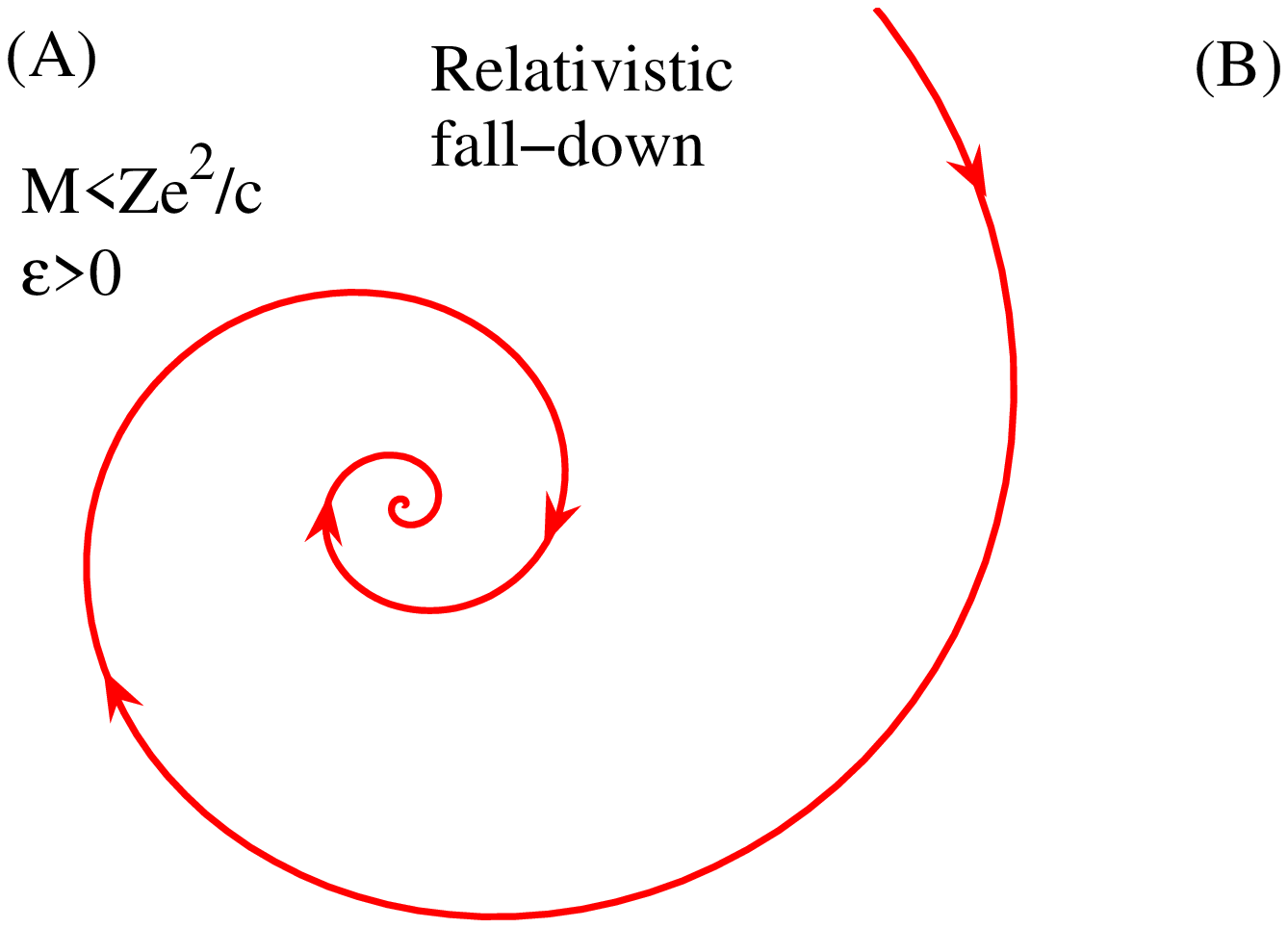}
\includegraphics[height=1.45in]{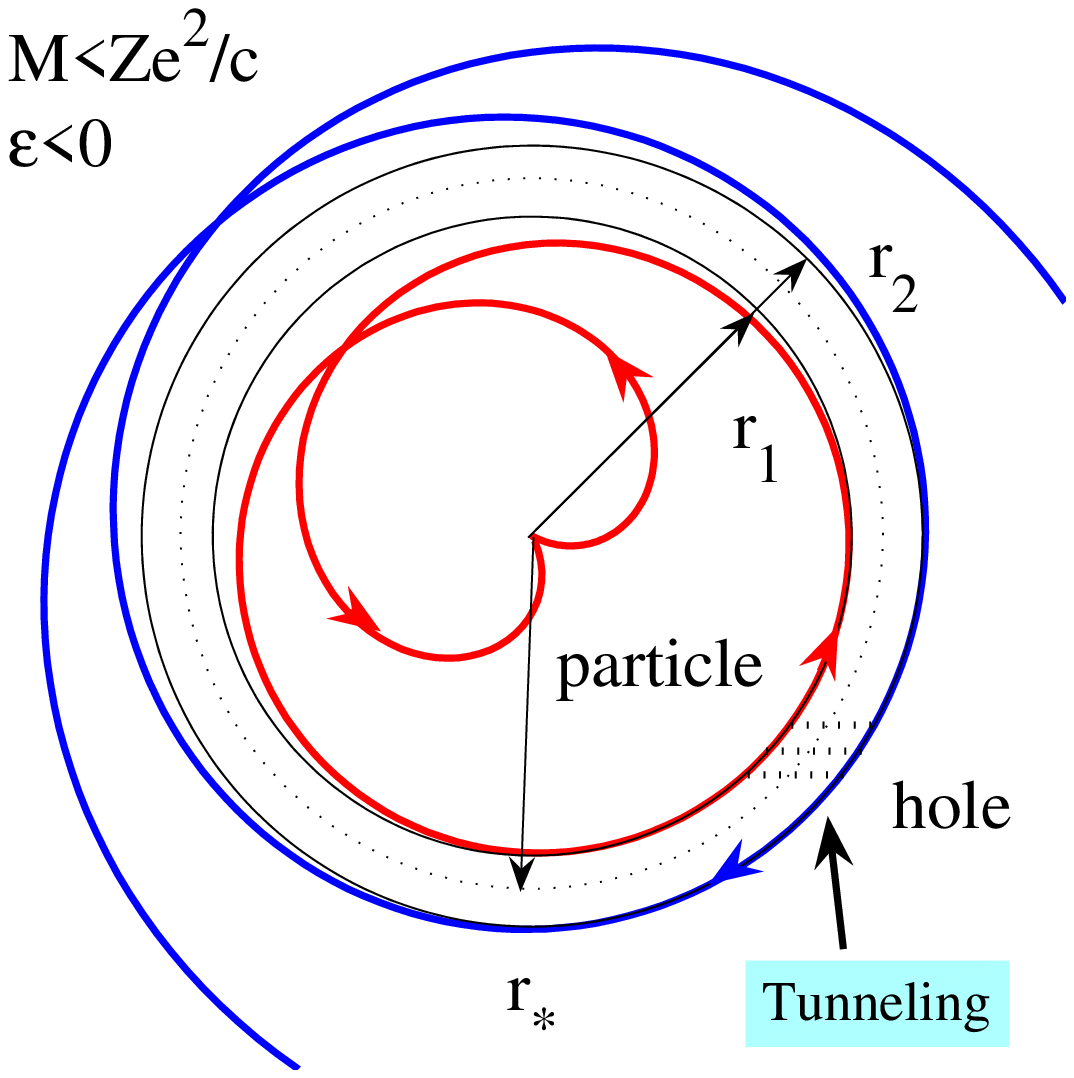}
\includegraphics[width=3.4in]{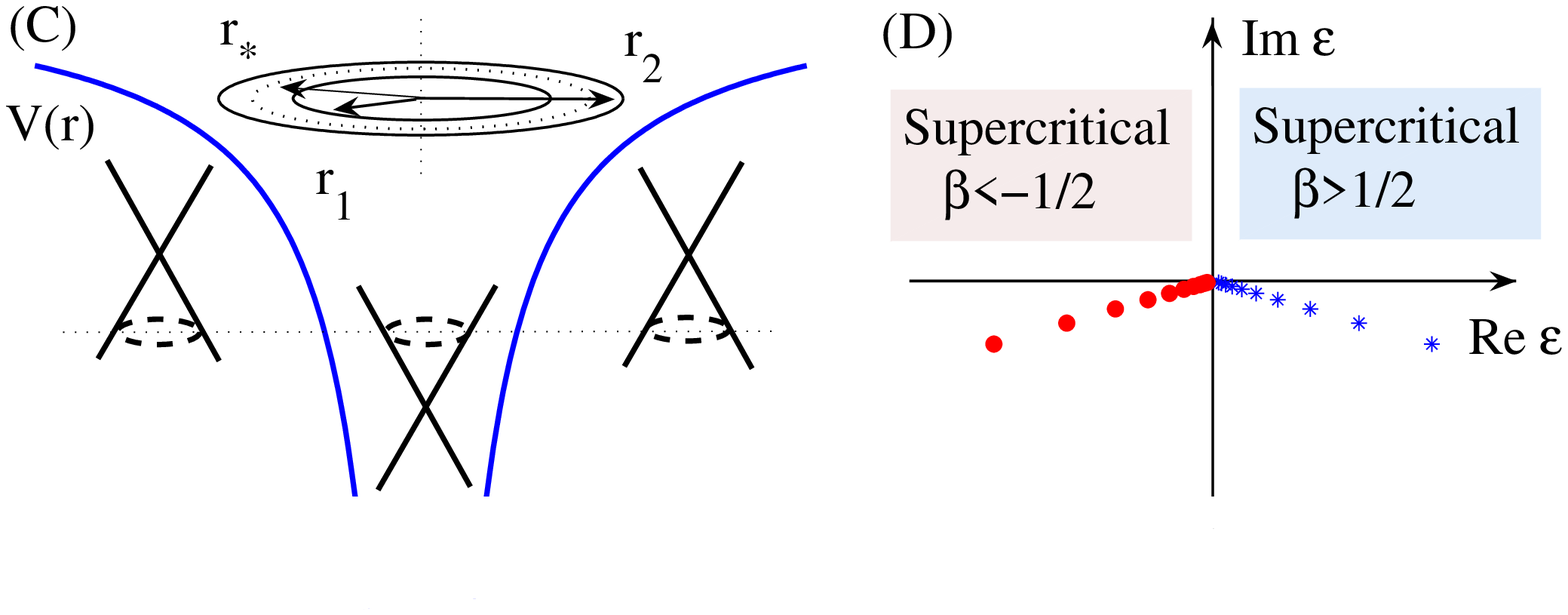}
\vspace{-0.85cm}
\caption[]{
Classical and quantum picture of atomic collapse
due to electron with angular momentum $M<M_c$
falling on the nucleus. Trajectories obtained from (\ref{eq:p_r})
for (A) positive and (B) negative energy $\epsilon$ of a massless Dirac particle
are shown.
(B,C) At $\epsilon<0$ there are collapsing particle trajectories and non-collapsing
hole trajectories, separated by a classically forbidden region, 
the annulus $r_1<r<r_2$
($r_{1,2}=r_\ast \mp Mv_F/\epsilon$, $r_\ast=Ze^2/|\epsilon|$).
Tunnel coupling to the continuum at $r>r_2$ defines a family of quasistationary states
with complex energy spectrum (D),
appearing abruptly when the potential strength 
exceeds the stability threshold $|\beta|=\frac12$.
}
\label{fig:schematic}
\end{figure}

Bohr's theory of an atom has explained that, while an electron 
is irresistibly pulled to the nucleus by the Coulomb force,
it is prevented from falling on it 
by the quantum mechanical zero-point motion.
This balance, however, becomes 
more delicate in the relativistic theory. The effects 
undermining the stability of matter
arise already in classical dynamics,
where electron trajectory can spiral around the 
nucleus and eventually fall down on it\,\cite{Darwin1913}
(see Fig.\ref{fig:schematic}a,b),
provided that
electron angular momentum
is small enough:
$M<M_c=Ze^2/c$, where $Z$ is nuclear charge. 
Quantum mechanics partially saves matter from collapse by imposing 
the angular momentum quantization $M=n\hbar$, which makes the relativistic 
fall-down possible only for heavy $Z>\hbar c/e^2\approx 137$.


Early work on the Dirac-Kepler problem has revealed bizarre 
properties of atoms with nuclear charge in excess of $Z=137$,
posing as a fundamental bound on the periodic table of elements extent 
at large $Z$.
The breakdown at
$Z>137$
of the low-$Z$ solution of the Dirac equation
requires accounting for a finite
nuclear radius\,\cite{Pomeranchuk}. 
The resulting electron states dive into the hole continuum
at $Z>170$ and decay by positron emission\,\cite{Zeldovich,SLAC,Greiner_book}.
These phenomena,
never observed in the laboratory due to the difficulty of 
producing heavy nuclei,
should be more readily accessible in graphene 
owing to its large ``fine structure constant,''
$\alpha = e^2 /\hbar v_F\approx 2.5$,
where $v_F\approx 10^6\,{\rm m/s}$ is the velocity of Dirac excitations.

Charge impurities are an essential ingredient of our current understanding
of transport in graphene.
Scattering on charge impurities explains\,\cite{Nomura,Ando,dassarma}
the linear density dependence
of conductivity in this material\,\cite{Novoselov05}, 
and is thus regarded as one of the main factors limiting carrier mobility.
Recent investigations of screening of impurity potential\,\cite{Mirlin,Shytov07,Pereira07,Biswas07,Fogler07} 
have only reinforced these conclusions,
making graphene an ideal test system for the theory
of Coulomb scattering\,\cite{Shytov07,Novikov07} of massless Dirac particles.

In this work we show that,
although massless particles cannot form bound states,
an infinite family of quasi-bound states appears abruptly when the Coulomb
potential strength exceeds a certain critical value $\beta=\frac12$.
These states are obtained from Bohr-Sommerfeld quantization
of {\it collapsing orbits} which descend on the point charge,
similar to how the hydrogenic Rydberg states are found from
circular orbits.
The energies of these states converge on zero, 
$\epsilon_n\to0$ at large $n$, whereas their radii diverge, similar
to the Rydberg states.
These results are corroborated by an exact solution
of the 2D Dirac-Kepler problem.
In graphene, the effective Coulomb potential strength\,\cite{Shytov07} is
given by $\beta=Ze^2/\kappa \hbar v_F$
with intrinsic dielectric constant $\kappa\approx 5$, 
and therefore the critical value $\beta=\frac12$
can be reached already for the impurity charge $Z\gtrsim 1$.
This is a lot more convenient 
from the experimental point of view than $Z>170$ in heavy atoms.

Coupling of quasi-Rydberg states to the Dirac continuum, 
mediated by Klein tunneling\,\cite{Greiner_book},
leads to strong resonances in the scattering cross-section, 
manifest in transport,
and to striking effects in local properties that can be probed by
deliberately introducing impurities with $Z\gtrsim 1$
in graphene. 
Univalent charge impurities, 
such as K, Na, or ${\rm NH_3}$, all commonly used in graphene,
are on the border of the supercritical regime. 
To investigate this regime experimentally,
one can use divalent or trivalent dopants such as alkaline-earth or rare-earth
metals. They are frequently used to prepare intercalated graphite compounds\cite{Dresselhaus05},
e.g., Ca and Yb \cite{dop1} ($Z=2$), La \cite{dop2} and Gd \cite{dop3} ($Z=3$).
Recently, spectroscopic experiments on graphene doped by Ca have been
reported\,\cite{dop4}.

Recent literature\,\cite{Mirlin,Shytov07,Pereira07,Biswas07,Fogler07} 
investigated the problem of screening of charge impurities,
which depends
on the polarization of the Dirac vacuum\,\cite{Shytov07}. 
This effect
is mostly inconsequential in atomic physics 
due to its short spatial scale set by the Compton wavelength 
$\lambda=h/mc\approx 2.4\times 10^{-3}\,{\rm nm}$. 
In the massless case of graphene, however,
it leads to long-range polarization,
appearing above
the critical value $\beta=\frac12$ of the impurity charge\,\cite{Shytov07}. 
These studies indicate that $\beta=\frac12$
separates two very different regimes of screening,
essentially perturbative at $\beta<\frac12$\,\cite{Mirlin,Biswas07}, 
and nonlinear at $\beta>\frac12$\,\cite{Shytov07,Pereira07,Fogler07}.

To explain 
why the quasi-bound states appear at large $\beta$, we
consider fermions with
energy $\epsilon<0$ in the potential
$V(r)=-Ze^2/r$. Since the kinetic energy $K=\epsilon-V(r)$
vanishes at $r_\ast=Ze^2/|\epsilon|$, the polarity of carriers
changes sign inside the disk $r<r_\ast$ (see Fig.\ref{fig:schematic}c). 
If $r_\ast$ exceeds particle
wavelength $\lambda=\hbar v_F/|\epsilon|$, 
which happens for
$\beta=Ze^2/\hbar v_F\gtrsim 1$,
quantum states 
can be trapped at $r\lesssim r_\ast$. 
These states will have finite lifetime due to
Klein tunneling through the barrier
at $r\approx r_\ast$.
Crucially, since the ratio $r_\ast/\lambda$
is independent of $\epsilon$,
this reasoning predicts infinitely many quasi-bound states (see Fig.\ref{fig:schematic}d).
These states can be constructed quasiclassically,
from relativistic dynamics described by the Hamiltonian
$H=v_F|\vec p|+V(r)$, where $V(r)=-Ze^2/r$.
The collapsing trajectories with angular momenta
$M< M_c=Ze^2/v_F$
are separated from non-falling trajectories by a centrifugal barrier.
This is manifest in the radial dynamics 
%
\begin{equation}
\label{eq:p_r}
p_r^2  = v_F^{-2}\left(\epsilon + \frac{Ze^2}{r}\right)^2 - \frac{M^2}{r^2}
, 
\end{equation}
where $p_r$ is the radial momentum.
This defines a classically forbidden region, the annulus $r_1<r<r_2$,
$r_{1,2}=(Ze^2 \mp Mv_F)/\epsilon$, where (\ref{eq:p_r}) is negative.
The quasi-bound states trapped by this barrier
can be found from the Bohr-Sommerfeld quantization condition
$\int_{r_0}^{r_1} p_r dr=\pi\hbar n$, where $r_0$ is the lattice
cutoff
(cf. Refs.\cite{Silvestrov07,Chen07}). 
Evaluating the integral with logarithmic accuracy, we obtain
$ \gamma \ln \frac{Ze^2}{r_0\epsilon}
= \pi\hbar n$, where $\gamma\equiv \lp M_c^2-M^2\rp^{1/2}$,
which gives the quasi-Rydberg states
%
\be \label{eq:En_quasiclass}
\epsilon_n \approx
\frac{Ze^2}{r_0}e^{-\pi \hbar n/\gamma}
,\quad
n>0
.
\ee
%
The energies (\ref{eq:En_quasiclass}) are equally spaced on the log scale with the separation diverging
as $1/\gamma$ at the threshold $M_c\approx M$.

To find the transparency of the barrier, 
we integrate $\Im p_r$ to obtain the tunneling action
%
\begin{equation}\label{eq:S_coulomb}
S = \int_{r_1}^{r_2} dr \sqrt{\frac{M^2}{r^2} - \left(\frac{\epsilon}{v_F} + \frac{M_c}{r}\right)^2}
= \pi \left(M_c - \gamma\right)
.
\end{equation}
%
Taken near the threshold
$\gamma\approx 0$, 
the transparency $e^{-2S/\hbar}$
gives the width
$\Gamma_n \sim |\epsilon_n|\exp (-2\pi Ze^2/\hbar v_F)$.
Notably, since $S$ has no energy dependence,
all the states (\ref{eq:En_quasiclass}) 
feature the same width-to-energy ratio.

\begin{figure}
\includegraphics[width=3.4in]{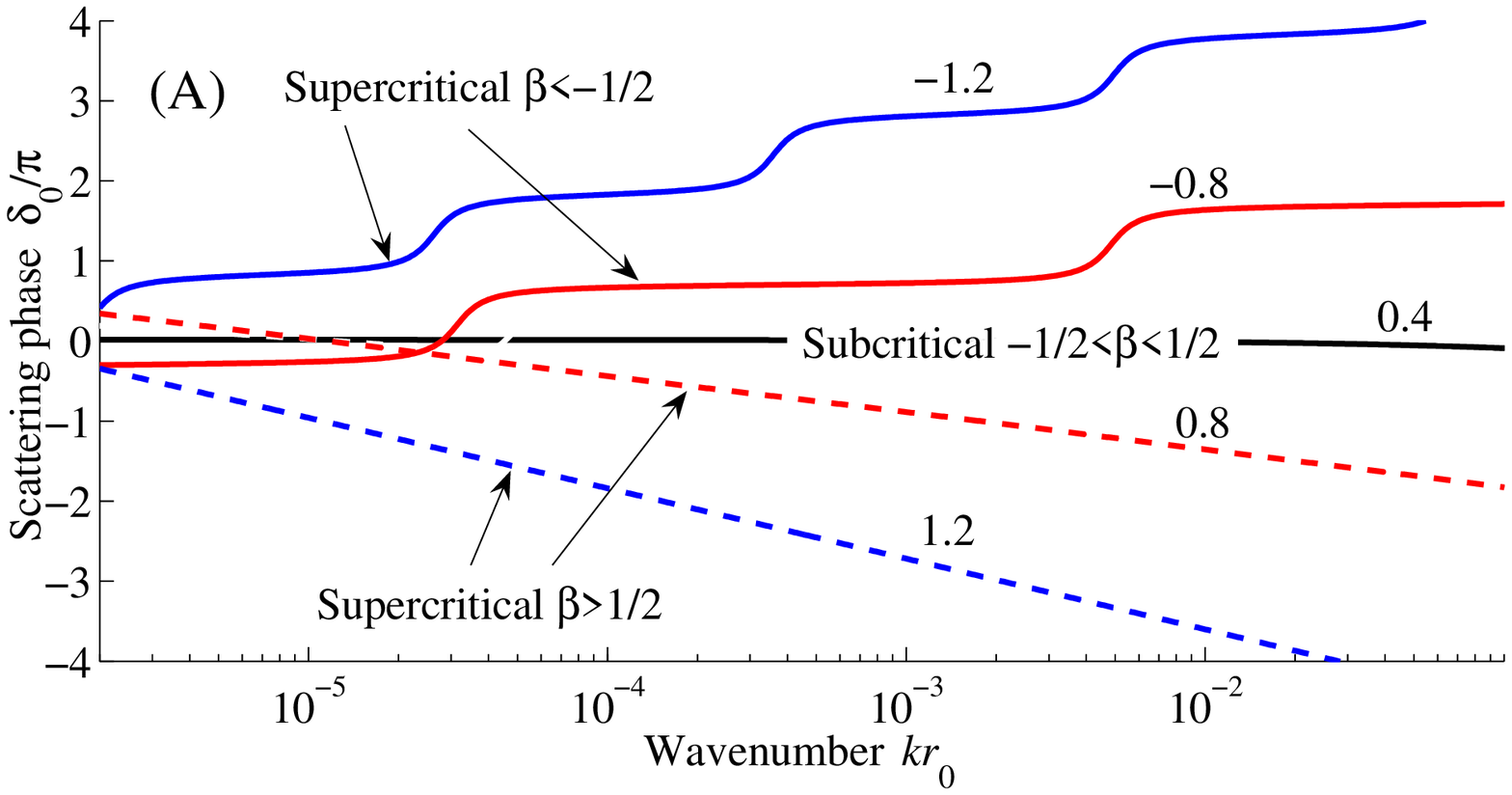}
\includegraphics[width=3.4in]{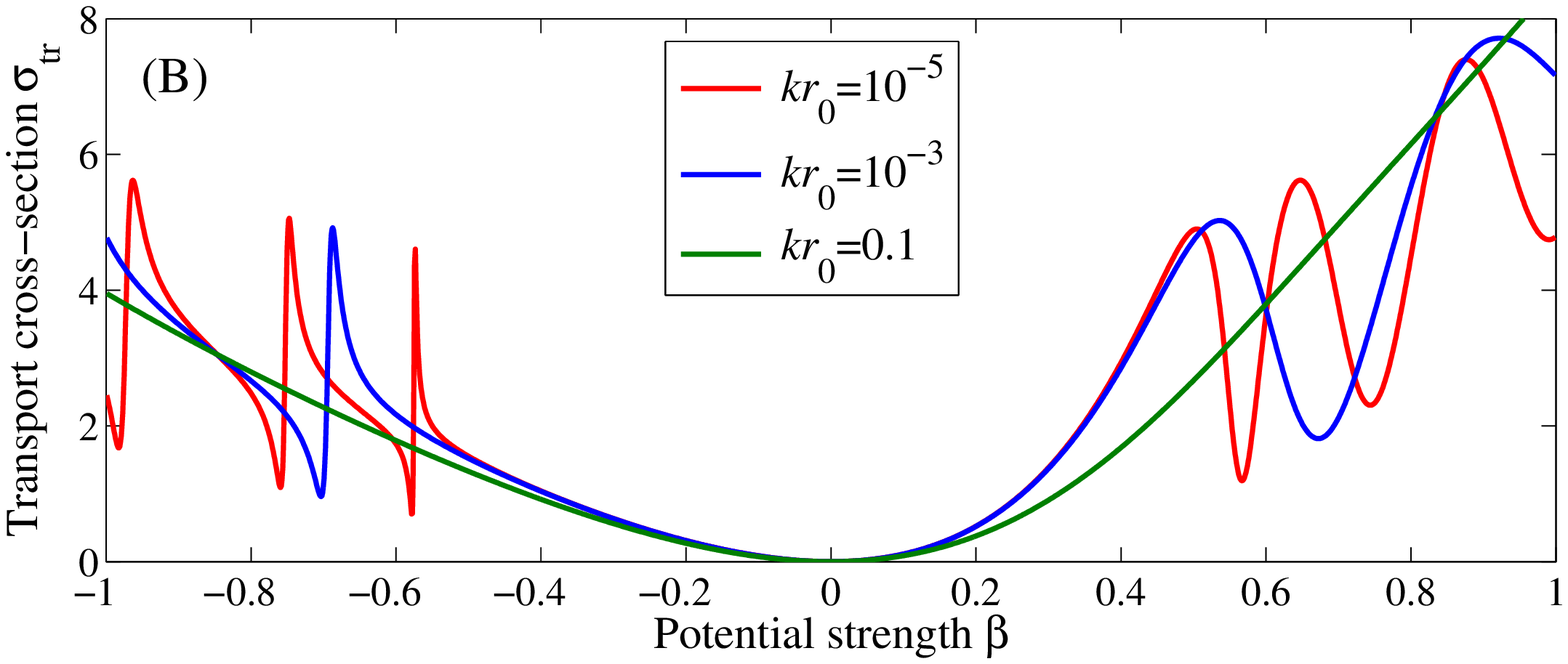}
\vspace{-0.85cm}
\caption[]{
(A) Scattering phase $\delta_0$
at negative energy $\epsilon=-v_Fk<0$.
The kinks correspond to the quasi-bound states
trapped by the impurity potential 
of supercritical strength $\beta<-1/2$,
as illustrated in Fig.\ref{fig:schematic}.
(B) Transport cross-section (\ref{sigma-transp-general}) {\it vs.} potential strength. 
Fano resonances corresponding to quasi-bound states occur at $\beta<-\frac12$.
The oscillatory behavior at $\beta>\frac12$
results from the energy dependence
$\delta_m(k)\sim -\gamma_m\ln kr_0$. 
The cross-section asymmetry upon $\beta\to -\beta$,
with the values at $\beta<0$ typically lower than at $\beta>0$,
reflects that
the Klein barrier prevents particles from reaching 
the region of strong scattering $r\sim 0$.
}
\label{fig:phases}
\end{figure}

It is instructive to compare these results to the exact solution 
of the Coulomb scattering problem. For that,
we consider the Dirac equation for a massless electron in 
a potential $V(r)=\frac{\beta}{r}$, where $\beta\equiv - Ze^2/\hbar v_F$.
(It will be convenient to include the minus sign in $\beta$ 
to explicitly account for attraction.)
Performing standard angular decomposition and solving the radial
equation separately in each angular momentum channel\,\cite{Shytov07}, 
one finds the scattering phases $\delta_m(k)$ that
behave differently in
the three regimes: 
(i) $\beta>\beta_m$, 
(ii) $-\beta_m<\beta<\beta_m$ 
and (iii) $\beta<-\beta_m$ ($\beta_m=|m+\frac12|$).
As illustrated in Fig.\ref{fig:phases}A, $\delta_m(k)$ are energy-independent
in case (ii) and have a logarithmic dependence $\delta_m(k)\sim -\gamma \ln kr_0$
in case (i), where $\gamma=\sqrt{\beta^2-\beta_m^2}$. In the case (iii) the
dependence is
described by kinks of height $\pi$ equally spaced 
on a log scale:
\be\label{eq:delta_exact}
e^{2i\delta_m(k)}= e^{\pi i\beta_m}
\frac{
z+ e^{i\chi(k)} 
}{
1 + e^{i\chi(k)} z^\ast
}
,
\ee
(see \cite{Shytov07}, Eq.(20)) where $z= \frac{e^{\pi \gamma}}{\eta} 
\frac{\Gamma (1 + 2 i \gamma)}{\Gamma (1 - 2 i \gamma)}
\frac{\Gamma (1 - i \gamma + i \beta)}{\Gamma (1 + i \gamma + i \beta)}$,
and
\be
\label{chi-result}
{\textstyle 
\chi (k) = 2 \gamma \ln 2kr_0 
  + 2 \arctan\frac{1+\eta}{1-\eta}
,\quad
\eta=\sqrt{\frac{\beta - \gamma}{\beta + \gamma}}
.
}
\ee
The average winding rate of the phase
in case (iii), $\bar\delta_m(k)\sim \gamma \ln kr_0$, 
is the same as in case (i) up to a sign (see Fig.\ref{fig:phases}).
The kinks 
signal the appearance of quasi-bound states at negative energies.


To find the quasi-bound states, we seek a  
scattering state with complex energy in which there is no incoming wave. 
This implies vanishing of the numerator (denominator) of (\ref{eq:delta_exact})
at $\epsilon<0$ ($\epsilon>0$).
In the $\epsilon<0$ case
we obtain an equation for $k$:
%
%
%
%
$e^{i \chi(k)} = -z$. The right-hand side of this equation 
in general has a non-unit modulus,
which makes it impossible to satisfy 
it by a real $k$.
Complex solutions of $\chi(k)=-i\ln(-z)-2\pi i n$
resemble those obtained quasiclassically, Eq.(\ref{eq:En_quasiclass}).

For a more direct comparison, let us consider $\beta$ near the threshold
$\beta=\beta_m$. Expanding in small $\gamma$, 
we find solutions
similar to our quasiclassical result 
(\ref{eq:En_quasiclass}),
\begin{equation}
\label{eq:k_n-exact}
k_n = \frac{c}{2 r_0}
\exp\lp -\frac{\pi}{\gamma}n  
- i\lambda \rp
,\quad
n>0
,
\end{equation}
where $\lambda=\frac{\pi}{1 - e^{-2\pi \beta}}$ and the prefactor $c$
is of order one. 
(We suppress $n\le0$, 
since Eq.(\ref{chi-result}) holds only for $kr_0\ll1$.)

The interpretation of the solutions (\ref{eq:k_n-exact}) depends on
the sign of $\beta$.
For near-critical negative values $\beta\approx -\frac12$
we have $\arg k_n\approx 0.045\pi$, such that $k_n$ have small imaginary parts,
defining sharp resonances of width
\begin{equation}
\label{qb-width}
\textstyle{\frac12} \Gamma = -\Im \epsilon =\textstyle{\frac{\pi}{e^\pi-1}}|\epsilon|
\approx  0.14\,|\epsilon|
.
\end{equation}
In contrast, $\arg k_n\approx -1.045\pi$ for positive $\beta\approx \frac12$, 
i.e. complex $k_n$'s 
are rotated by more than $180^{\rm o}$ away from positive 
semi-axis.
Thus there are no long-lived states with positive $k$ ($\epsilon<0$). 
Instead, since $\arg k_n\approx \pi$, in this case 
all $k_n$'s are found near the negative real semi-axis 
($\epsilon>0$). 
This is in agreement with particle/hole symmetry.



Before discussing manifestations in graphene,
where at finite carrier density the $1/r$ potential is screened, 
we note that the essential physics
will be unaffected by screening 
as long as the quasi-bound states persist. At finite density,
the RPA screening length is comparable to the Fermi wavelength 
$\lambda_F=\hbar v_F/\epsilon_F$\,\cite{Nomura,Ando,dassarma}, 
whereas our quasiclassical estimate of the state radius
gives $r_1\approx (M-M_c)v_F/|\epsilon|$.
The latter is much smaller than $\lambda_F$ near $\beta=\frac12$,
which means that RPA screening is non-detrimental for these states.
Similarly, estimates for nonlinear screening\,\cite{Shytov07}
indicate that its effect is inessential at weak coupling,
leaving enough room for quasi-bound states.

\begin{figure}
\includegraphics[width=3.4in]{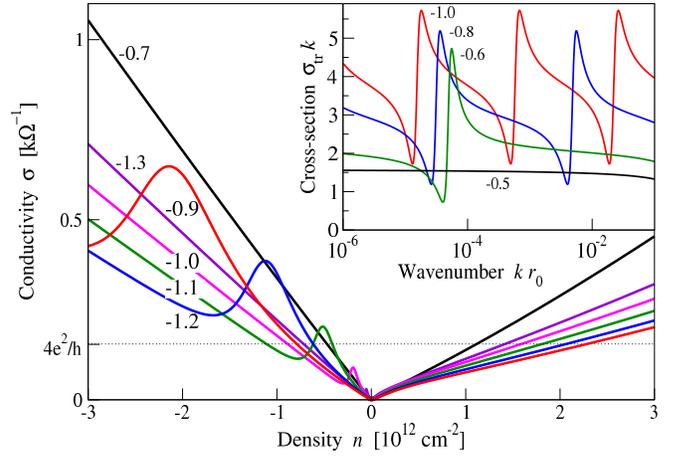}
\vspace{-0.45cm}
\caption[]{
Peak structure in the ohmic conductivity 
(\ref{eq:Drude}),(\ref{sigma-transp-general})
for overcritical $\beta$ 
occurring when the Fermi energy is aligned with resonances.
The values of $\beta$ are indicated near each trace.
(Parameters used: $n_{\rm imp}=3\cdot 10^{11}\,{\rm cm}^{-2}$,
$r_0=0.25\,{\rm nm}$)
Inset:
Fano resonance structure 
in the cross-section (\ref{sigma-transp-general})
at negative energies 
with the overall $1/k$ dependence factored out.
}
\label{fig:sigma-transport}
\end{figure}

Resonance scattering on the quasi-bound states will manifest itself
in the dependence of transport properties on the carrier density.
Here we analyze electrical conductivity described by the Drude-like model
(see Ref.\cite{Nomura}):
\be\label{eq:Drude}
\sigma=\frac{e^2}{h}2\epsilon_F\tau
,\quad
\tau^{-1}=v_F n_{\rm imp}\sigma_{\rm tr}
,
\ee
where $\epsilon_F$ is the Fermi energy, 
$n_{\rm imp}$ is the concentration of charge impurities 
and $\sigma_{\rm tr}$
is the transport scattering cross-section for one impurity.
We use the 2D scattering
amplitude partial wave decomposition  
\begin{equation}
f(\phi) = \frac{2i}{\sqrt{2\pi i k}} \sum_{m = 0}^{\infty} 
   (e^{2i \delta_m} - 1) \cos(m + \textstyle{\frac12}) \phi
, 
\end{equation}
(see Ref.\cite{KatsnelsonSSC}, Eq.(47)) to evaluate transport cross-section
%
\begin{equation}
\label{sigma-transp-general}
\sigma_{\rm tr} = \int d\phi (1 - \cos\phi) |f(\phi)|^2 = \frac{4}{k} \sum_{m=0}^{\infty} \sin^2 \theta_m
, 
\end{equation}
$\theta_m=\delta_{m} - \delta_{m + 1}$, 
with the phases $\delta_m$  given by (\ref{eq:delta_exact})
for overcritical channels (see Ref.\cite{Novikov07} for subcritical channels).

For subcritical potential strength 
the phases are energy-independent 
and thus $\sigma_{\rm tr}$
scales as $1/|\epsilon|$,
giving conductivity (\ref{eq:Drude})  
linear in the carrier density\,\cite{Nomura,Ando,dassarma}.
For $|\beta|>\frac12$ the contribution of the subcritical
channels still scales as $1/|\epsilon|$, while the overcritical channels, 
because of energy-dependent $\delta_m(k)$, 
give an oscillatory contribution 
({\it cf.} Fig.\ref{fig:phases}B). 
These oscillations, shown in Fig.\ref{fig:sigma-transport} inset, 
have a characteristic form of Fano
resonances centered at $\epsilon_n$.
In this regime the conductivity (\ref{eq:Drude}) 
exhibits peaks at the densities 
for which the Fermi energy $\epsilon_F$ alignes with $\epsilon_n$.
As evident from Fig.\ref{fig:sigma-transport}, 
the peak position 
is highly sensitive to the potential strength $\beta$,
changing by an order of magnitude when $\beta$ varies from
$-1.0$ to $-1.3$, which is a combined effect of $\epsilon_F$
quadratic dependence on density
and of the exponential
dependence in (\ref{eq:k_n-exact}).



Another striking feature in the conductivity {\it vs.} density plots
in Fig.\ref{fig:sigma-transport} is the $n\to -n$ asymmetry,
which results from the scattering cross-section being typically lower
at $\epsilon<0$ than at $\epsilon>0$. Such asymmetry, noted already
in the subcritical regime\,\cite{Novikov07}, becomes more prominent
in the supercritical regime
because of the Klein barrier preventing particles with negative energies
from reaching the strong scattering region $r\sim0$.

\begin{figure}
\includegraphics[height=1.72in]{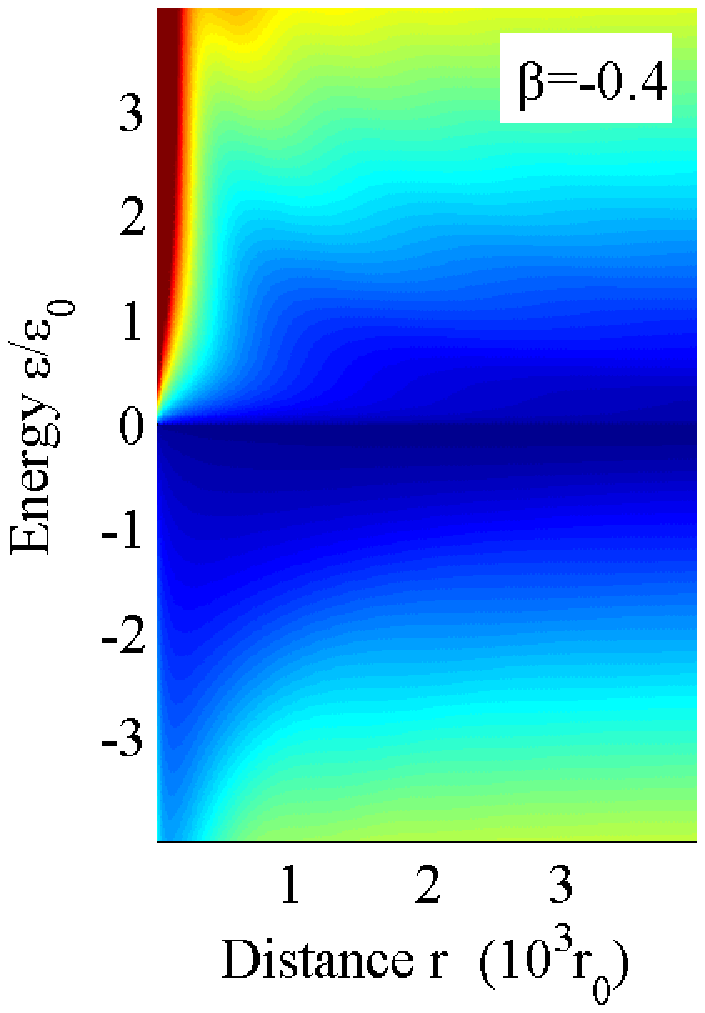} 
\includegraphics[height=1.72in]{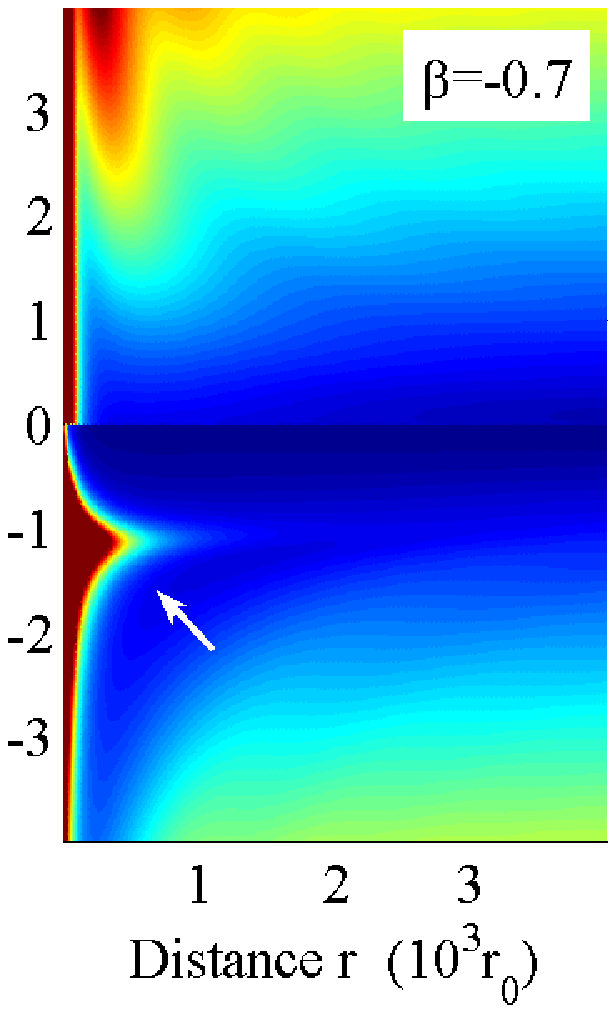} 
\includegraphics[height=1.72in]{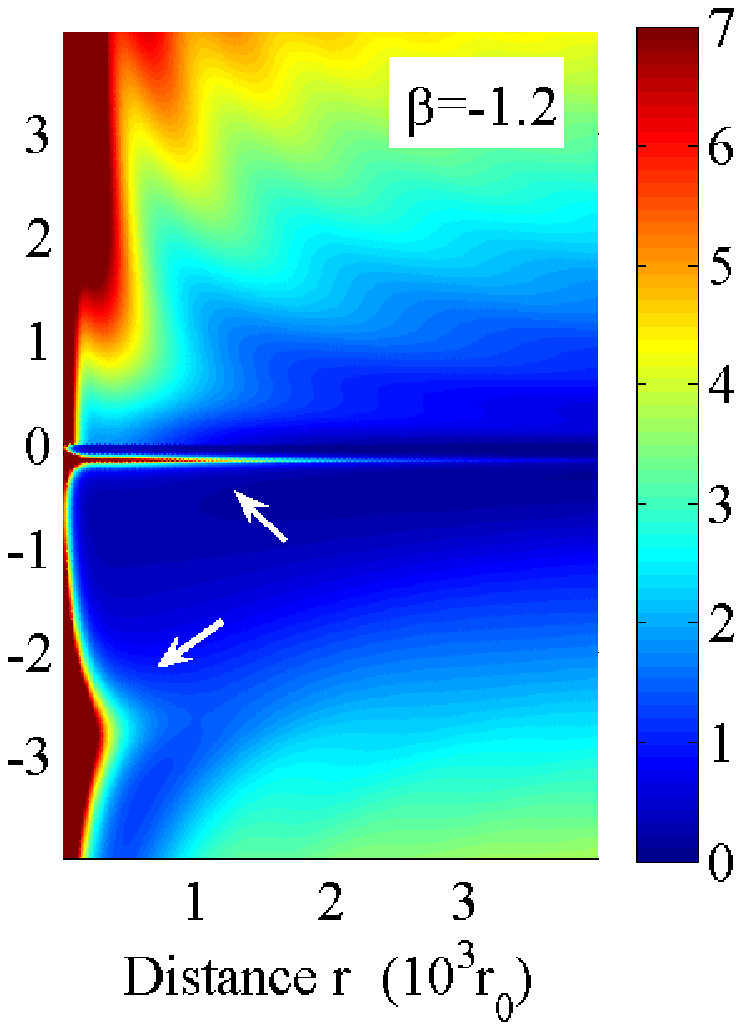} 
\vspace{-0.5cm}
\caption[]{
Spatial map of the local density of states (LDOS) 
near a charge impurity, Eq.(\ref{nu-general}). 
The signatures of the quasi-bound states are resonances 
appearing at $\beta<-\frac12$ at $r\sim 0$ and $\epsilon<0$ (marked by arrows).
Note the localization length that scales inversely with $\epsilon$,
and the linewidth proportional to $\epsilon$, 
as predicted by Eq.(\ref{qb-width}).
The intensity of the resonances is well in excess of 
the asymptotic value $\nu(\epsilon) \propto |\epsilon|$ at large $r$.
Periodic modulation at $\epsilon>0$ with maxima at $kr\approx \pi n$ 
is the standing
wave oscillation\,\cite{Shytov07}
($k=\epsilon/\hbar v_F$, $\epsilon_0=10^{-3}\hbar v_F/r_0$).
}
\label{fig:LDOS}
\end{figure}

The signatures of quasi-bound states,
similar to those in conductivity, will be featured by other transport
coefficients. In particular, they will be strong in the thermoelectric
response because it is proportional to the energy derivative of $\sigma_{\rm tr}$.
Yet the most direct way to observe these states is via
the local density of states (LDOS)
\begin{equation}
\label{nu-general}
\nu (\epsilon, r) 
= \frac{4}{\pi \hbar v_F}\sum_m 
|\psi (k_\epsilon, r)|^2
,
\quad
k_\epsilon = - \frac{\epsilon}{\hbar v_F}
,
\end{equation}
where 
$\psi$ is the two-component Dirac wave function (5)\,\cite{Shytov07}.
This quantity can be directly measured 
by scanning tunneling spectroscopy probes.
Evaluating the sum over $m$ in (\ref{nu-general}),
we obtain LDOS map shown in Fig.\ref{fig:LDOS}.

Several quasi-bound states are seen
in LDOS maps (Fig.\ref{fig:LDOS}) as local resonances at $\epsilon<0$. 
The values of $\beta$ 
were chosen to illustrate
that the width $\Gamma$ of each resonance scales with $\epsilon$,
while its spatial extent scales as $1/\epsilon$, in agreement with
our quasiclassical analysis and Eq.(\ref{qb-width}).

A distinct advantage of local probes,
as opposed to transport,
is that the supercritical impurities do not need to be a majority. 
In fact, it suffices
to locate just one non-univalent impurity and perform STM imaging
in vicinity. Alternatively, one can identify groups of two or three
univalent impurities
that together will act as one supercritical impurity,
or even deliberately create such a group by inducing
local charge by voltage applied to STM tip.

The resonances (\ref{eq:k_n-exact}) 
also give rise to anomalously strong Kondo-like effects.
The striking property of the linewidth (\ref{qb-width}),
namely, its proportionality
to the energy (see Fig.\ref{fig:LDOS}),
indicates that the dwell time diverges at $\epsilon\to0$.
This divergence compensates the suppression of Kondo temperature
by the Dirac $|\epsilon|$ density of states.
Standard estimates\,\cite{Hewson_book} for the Anderson model 
with the localized spin state
associated with one of our resonant levels yield the Kondo
temperature exponent that exhibits no suppression at $\epsilon_F\to0$.
This is in contrast to 
the Kondo problem
with extraneous spin impurities\,\cite{Hentschel,Baskaran}.


In summary, although massless particles
are incapable of forming discrete states, 
an infinite family of quasi-Rydberg states can 
appear in a Coulomb potential of supercritical strength.
These quasi-bound states manifest themselves in a variety
of physical properties, in particular in resonant scattering and local 
resonances, providing a striking signature of the atomic collapse regime
that can be modeled using charge impurities in graphene.

This work is supported by the DOE
(contract DEAC 02-98 CH 10886),
FOM (The Netherlands), NSF MRSEC (DMR 02132802) and
NSF-NIRT DMR-0304019.

\end{document}